\documentclass[notitlepage,prl,letter,reprint,footinbib]{revtex4-1}

\usepackage{graphicx}
\usepackage{amsmath}
\usepackage{amssymb}
\usepackage{hyperref}
\usepackage{tikz}
\usepackage[english]{babel}

\newcommand{\be}{\begin{eqnarray*}}
\newcommand{\ee}{\end{eqnarray*}}
\newcommand{\beq}{\begin{eqnarray}}
\newcommand{\eeq}{\end{eqnarray}}
\newcommand{\bequ}{\begin{equation}}
\newcommand{\eequ}{\end{equation}}

\newcommand{\br}{{\mathbf{r}}}

\newcommand{\bb}{{\mathbf{b}}}

\newcommand{\bv}{{\mathbf{v}}}

\newcommand{\dd}{\mathrm{d}}

\newcommand{\bx}{\hat{{\mathbf{x}}}}
\newcommand{\by}{\hat{{\mathbf{y}}}}

\newcommand{\ph}{{\phantom{\dagger}}}
\newcommand{\ket}[1]{\left|{#1}\right\rangle}
\newcommand{\bra}[1]{\left\langle{#1}\right|}

\newcommand{\zbb}{\mathbb{Z}}

\newcommand{\id}{\mathbb{I}}

\begin{document}
\title{Floquet Topological Order in Interacting Systems of Bosons and Fermions}
\author{Fenner Harper}
\author{Rahul Roy}
\affiliation{Department of Physics and Astronomy, University of California, Los Angeles, California 90095, USA}
\date{\today}
\begin{abstract}
Periodically driven noninteracting systems may exhibit anomalous chiral edge modes, despite hosting bands with trivial topology. We find that these drives have surprising many-body analogs, corresponding to class A, which exhibit anomalous charge and information transport at the boundary. Drives of this form are applicable to generic systems of bosons, fermions, and spins, and may be characterized by the anomalous unitary operator that acts at the edge of an open system. We find that these operators are robust to all local perturbations and may be classified by a pair of coprime integers. This defines a notion of dynamical topological order that may be applied to general time-dependent systems, including many-body localized phases or time crystals.
\end{abstract}
\maketitle
\textit{Introduction.}---
Time-dependent quantum systems can support a host of novel phenomena that are impossible to realize with a static Hamiltonian. These include topological adiabatic cycles \cite{Thouless:1983vd,Fu:2006jk,Teo:2010ji,Roy:2011uv,Zhang:2014dv,Lopes:2016ga}, Floquet analogs of topological insulators \cite{Kitagawa:2010bu,Jiang:2011cw,Rudner:2013bg,Thakurathi:2013dt,Asboth:2014bg,Nathan:2015bi,Titum:2015fl,Carpentier:2015dn,Fruchart:2016hk,Roy:2016via,Titum:2016km,Budich:2016tc,Roy:2016wo}, novel examples of driven symmetry-protected topological phases (SPTs) \cite{Khemani:2016gd,vonKeyserlingk:2016ea,Else:2016ja,Potter:2016cr,Roy:2016ka}, and phases which exhibit spontaneous symmetry breaking in the time domain, dubbed time crystals or $\pi$-spin glasses \cite{Khemani:2016gd,vonKeyserlingk:2016bq,Else:2016gf,vonKeyserlingk:2016ev,Yao:2017bu}. In addition to being of theoretical interest, much progress has been made towards realizing Floquet systems in the laboratory \cite{Kitagawa:2012gl,Rechtsman:2013fe,Fregoso:2013di,Wang:2013fe,Jotzu:2015kz,JimenezGarcia:2015kd,Maczewsky:2016fj}. 

Many of these unusual Floquet phases are distinguished by their anomalous edge behavior: while a periodic drive may have no overall effect on a closed system, its action at a boundary can be nontrivial. This is a kind of holography that signifies the presence of an inherently dynamical type of order. In this Letter, we introduce a set of 2D drives that generate dynamical topological order of this form in generic systems of interacting bosons, fermions, or spins. The topological order manifests as robust chiral edge modes at the boundary of an open system, which are stable to all perturbations and which cannot be generated by a 1D Hamiltonian. 

We study these drives by considering the action of the unitary evolution restricted to the edge of the system, finding that it may be classified by a pair of coprime integers. Through homotopy arguments, this defines a robust topological invariant that may be applied very generally to classify Floquet many-body phases (for example, by incorporating many-body localization (MBL) \footnote{See Supplemental Material, which includes Refs.~\cite{Nandkishore:2015kt,Bahri:2015ib,Potter:2015vna,Slagle:2015uo,Chandran:2014dk,Bauer:2013jw,Huse:2013bw,Zhang:2016hy,Zhang:2016vt,Abanin:2015bc,Lazarides:2015jd,Ponte:2015dc,Ponte:2015hm,Abanin:2016eva,Lazarides:2014ie,DAlessio:2013fv,Chandran:2014dk,Bahri:2015ib,DeRoeck:2016us}, for further details}). However, our approach may also be used to provide a topological classification of more exotic unitary evolutions, including those corresponding to time crystals or those with only partial MBL.

As motivation, we recall that two fundamental examples of SPT phases are those of class D, protected by particle-hole symmetry, and class A, protected by U(1) charge conservation. In the time-dependent case, a 1D class D system has a $\zbb_2\times \zbb_2$ classification \cite{Roy:2016via,Jiang:2011cw,Thakurathi:2013dt}, which persists in the presence of interactions \cite{vonKeyserlingk:2016ea,Potter:2016cr,Roy:2016ka}. In 2D, class A corresponds to the integer quantum Hall effect (IQHE) \cite{prange1987quantum}, which has a well-defined (static) integer classification that also persists in the presence of interactions \cite{Niu:1984uz}. A nontrivial driven system belonging to class A (without interactions) was given in Ref.~\onlinecite{Rudner:2013bg}, which shows IQHE-like chiral edge modes at the boundary of a 2D lattice, even though the bulk band has a Chern number of zero. This model is stable to disorder \cite{Titum:2016km} and has recently been realized using photonic waveguides \cite{Maczewsky:2016fj}.
 
In the first section of this Letter we examine the effect of interactions on this drive by constructing its many-body analogs, taking into account the particle statistics. We consider the anomalous action of the drive at the edge, and find that charge conservation protects this against any local charge-conserving 1D perturbation. The structure of these particle-based models motivates a more general set of exchange models, which we introduce in the context of spin systems. We find that these also exhibit a robust edge action, and we provide a classification scheme for their anomalous behavior.

\textit{Interacting Class A Drive.}---
We first introduce a unitary drive in class A that reduces to the drive of Ref.~\onlinecite{Rudner:2013bg} for a single-particle system. We recall that the unitary evolution operator for a Hamiltonian $H(t)$ is $U(T)=\mathcal{T}\exp\left(-i\int_0^T H(t)\dd t\right)$, where $\mathcal{T}$ is the time-ordering operator, and $T$ is the period of evolution. We are specifically interested in unitary loops, which we define to be an evolution which, in a closed system, satisfies $U(T)=\id$. In the corresponding open system, however, $U(T)$ will not necessarily be proportional to the identity. The component of $U(T)$ that acts in the vicinity of the boundary, which we call the effective edge unitary, characterizes the anomalous edge action of the evolution. Although this class of unitaries may seem somewhat restrictive, a classification of unitary loops in fact gives a very general classification of dynamical topological order in the space of unitary evolutions (see Supplemental Material \cite{Note1} and Ref.~\cite{Roy:2016ka}).

\begin{figure}[t]
\includegraphics[scale=0.29]{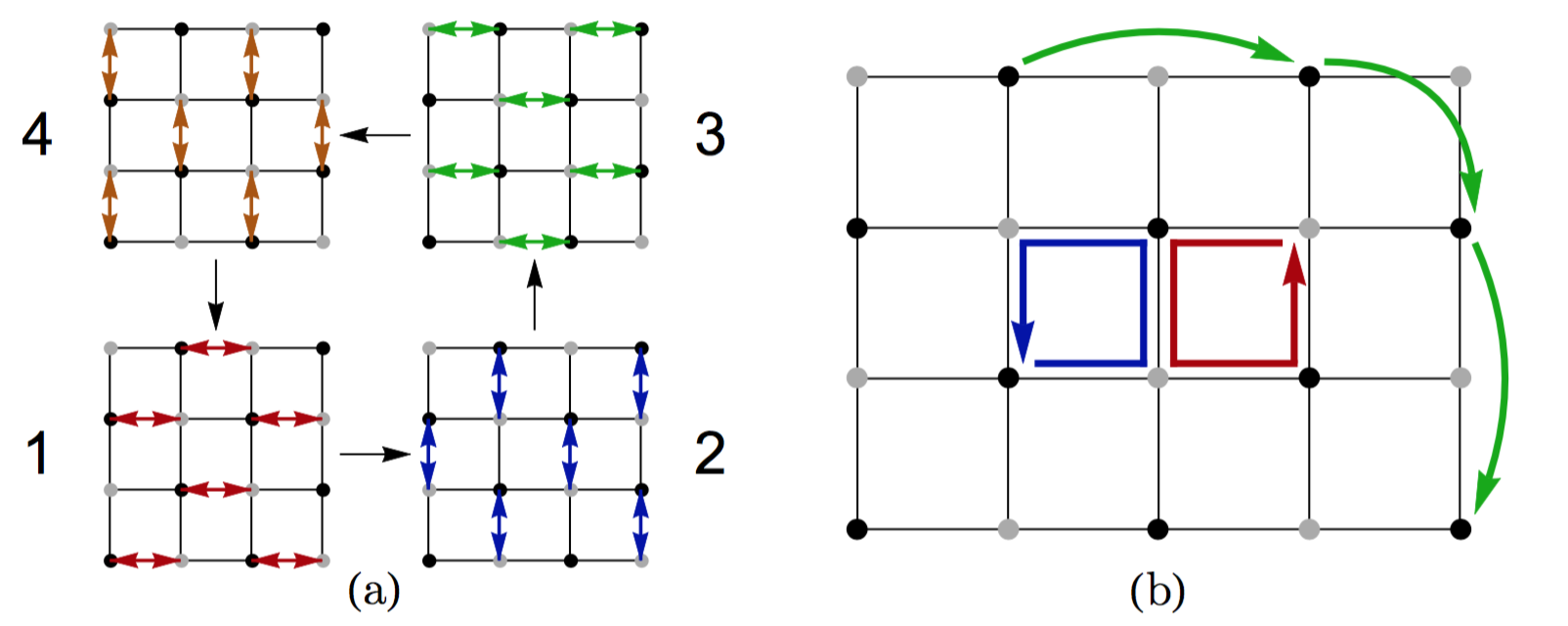}
\caption{(a) Four steps in the anomalous Floquet drives considered in the main text, based on the model of Ref.~\onlinecite{Rudner:2013bg}. (b) Representation of the action of the complete unitary drive. See main text for details.\label{fig:unitary_drive}}
\end{figure}

The drive in Ref.~\onlinecite{Rudner:2013bg} consists of four principal steps, each of which generates hopping across a different set of neighboring bonds, as shown in Fig.~\ref{fig:unitary_drive}(a). After the complete drive, a particle in the bulk returns to its initial position, but a particle located at certain positions on the edge is transported along the boundary, represented pictorially in Fig.~\ref{fig:unitary_drive}(b). For simplicity, we initially work with hardcore bosons. In this case, the unitary that generates a hop between two sites is
\begin{equation}
U^{B}_{\br\br'}=1+b^\dagger_{\br'}b^\ph_\br+b^\dagger_\br b^\ph_{\br'}-b^\dagger_{\br}b^\ph_{\br}-b^\dagger_{\br'}b^\ph_{\br'}+2b^\dagger_{\br'}b^\ph_{\br'}b^\dagger_{\br}b^\ph_{\br},\label{eq:2site_boson_unitary}
\end{equation}
where $b^\dagger_{\br}$ creates a boson on site $\br$ and satisfies $[b^\ph_\br,b_{\br'}^\dagger]=\delta_{\br,\br'}$ and $\left(b^\dagger_\br\right)^2=0$. It may be verified that this operator is unitary and that it acts on a general two-site state as 
\beq
U^{B}_{\br\br'} \left(b^\dagger_{\br^{\phantom{\prime}}}\right)^{n_\br}\left(b^\dagger_{\br'}\right)^{n_{\br'}}\ket{0}=\left(b^\dagger_{\br^{\phantom{\prime}}}\right)^{n_{\br'}}\left(b^\dagger_{\br'}\right)^{n_{\br}}\ket{0},
\eeq
with $n_{\br},n_{\br'}\in\{0,1\}$. Labeling the two sublattices as $A$ and $B$ (filled and open circles, respectively, in Fig.~\ref{fig:unitary_drive}), and setting the intersite spacing to one, each step of the unitary drive may be written $U^B_j=\prod_{\br\in A} U^{B}_{\br,\br+\bb_j}$, with $\bb_1=-\bb_3=(1,0)$ and $\bb_2=-\bb_4=(0,-1)$. The complete unitary drive is then $U^B=U^B_4U^B_3U^B_2U^B_1$, which can be written as the product of evolutions by four local Hamiltonians.

Within each step of the drive, the two-site operators $U^{B}_{\br,\br'}$ act on disjoint pairs of sites and commute. By tracking the position of a particular particle across all steps of the unitary, it can be verified that the action of the complete drive translates particles as in Fig.~\ref{fig:unitary_drive}(b). On a many-body product state, the unitary acts as a permutation of particle occupation numbers at the edge. Since the unitary acts identically on any product state, the permutation is also well defined for superposition states.

The effective edge unitary of the drive may be read off directly from the complete time evolution operator. Writing a generic many-body product state as $\ket{n_1,n_2,\ldots }$, where $n_j$ gives the boson occupation number on site $j$, the unitary drive maps product states onto product states through the relation $\ket{n_1',n_2',\ldots}=U^{B,\{n'\}}_{\{n\}}\ket{n_1,n_2,\ldots}$. From the discussion above, the matrix elements are 
\beq
U^{B,\{n'\}}_{\{n\}}=\prod_{j\in\mathrm{bulk}}\delta_{n_j,n_j'}\prod_{j\in\mathrm{edge}}\delta_{n_j,n_{j+1}'},
\eeq
where the sites at the edge have been indexed appropriately \footnote{We include states that lie on the physical edge but which are unchanged after the drive as part of the bulk.}. The effective edge unitary $U^B_{\rm eff}$ is characterized by the matrix elements of the second factor. 

It is natural to ask whether this many-body generalization applies also to fermions. We define fermionic unitary operators, $U^F_{\br\br'}$, $U^F_j$ and $U^F$ by replacing $b^\dagger_\br$ with $f^\dagger_\br$ in the bosonic definitions above. The operators $f^\dagger_\br$ satisfy $\{f^\ph_\br,f^\dagger_{\br'}\}=\delta_{\br,\br'}$ and have the occupation number-exchanging property 
\beq
U^{F}_{\br\br'} \left(f^\dagger_{\br^{\phantom{\prime}}}\right)^{n_\br}\left(f^\dagger_{\br'}\right)^{n_{\br'}}\ket{0}=\left(f^\dagger_{\br^{\phantom{\prime}}}\right)^{n_{\br'}}\left(f^\dagger_{\br'}\right)^{n_{\br}}\ket{0},
\eeq
with $n_{\br},n_{\br'}\in\{0,1\}$. In this case, the presence of the vacuum state $\ket{0}$ is important. For a many-body Slater determinant, anticommuting the relevant fermion operators so that they are adjacent to the vacuum will introduce an overall sign, which depends on the occupation of other lattice sites. In this way, the fermionic drive $U^F$, acting on a closed system, may return a Slater determinant state to itself or to minus itself; in an open system, the unitary translates particles at the edge only up to a sign. The fermionic matrix elements are related to their bosonic counterparts through $U^{F,\{n'\}}_{\{n\}}=\left(-1\right)^{s}U^{B,\{n'\}}_{\{n\}}$, where $s$ is an integer that depends nonlocally on $\{n'\}$ and $\{n\}$. For a superposition state, the unitary may introduce different signs for different components.

Nevertheless, the fermionic Floquet drive $U^F$ has many interesting properties and also exhibits anomalous edge behavior. Any charge distribution at the edge of a many-body state will be translated around the boundary. Furthermore, if the drive is run twice (which we call the ``doubled fermion drive''), then its action in the bulk is exactly the identity, since the sign factors square to one. This is reminiscent of fermionic Hamiltonians that avoid the sign problem. In this case, the bulk and edge behavior can be disentangled, and an effective edge unitary can be defined~\cite{Note1}.

\begin{figure}[t!]
\includegraphics[scale=0.27]{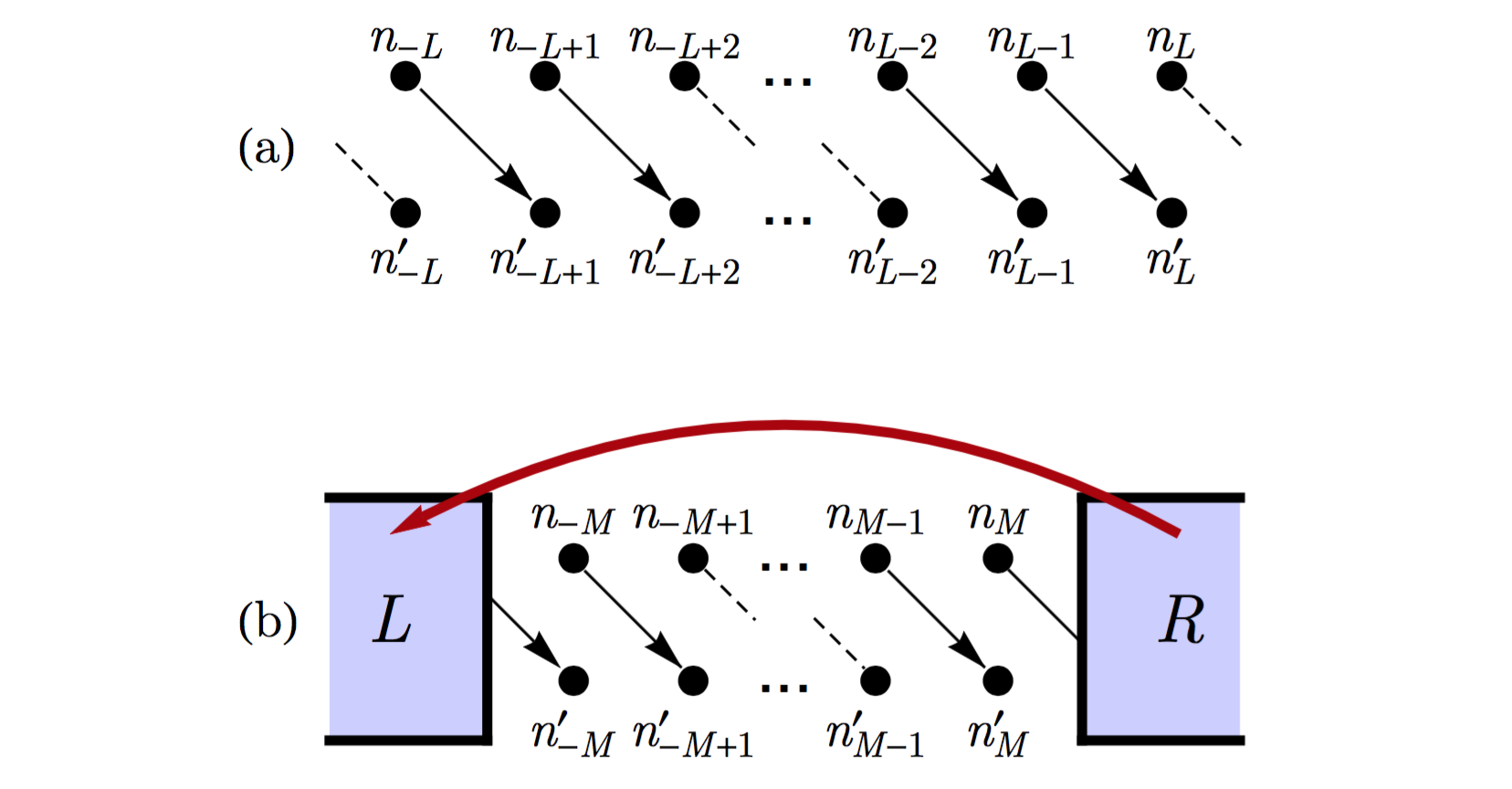}
\caption{(a) Permutation of boson occupation numbers at the edge under the anomalous Floquet drive. (b) Action of the putative open 1D unitary $\tilde{U}^B_{\rm op}$. The regions marked $L$ and $R$ are within a distance $v_{LR}T$ of the cut. See main text for details. \label{fig:edge_cartoon}}
\end{figure}

The drives described above have been constructed to give the desired edge behavior, and one might ask whether they are truly representative of a finite parameter space. We now argue that this is the case, and that the anomalous action is stable to local unitaries at the edge. We initially consider the bosonic version of the drive. To proceed, we consider the action of the effective unitary restricted to the 1D edge, which we take to have length $2L$. Acting on a product state, we find $U^B_{\rm eff}\ket{n_{-L},n_{-L+1},\ldots , n_{L}}=\ket{n_{L},n_{-L},\ldots, n_{L-1}}$, shown pictorially in Fig.~\ref{fig:edge_cartoon}(a). We will now assume that this unitary may be generated by a local 1D charge-conserving Hamiltonian $H(t)$ that acts for a finite time $T$ at the edge, and we will show that this leads to a contradiction \footnote{Throughout this document, a `local Hamiltonian' is made up of terms whose magnitudes are bounded by an envelope function that decays exponentially with the separation of sites involved. A `local unitary' is a unitary evolution derived from a local Hamiltonian.}. 

Since $H(t)$ is local, the complete drive has a maximum Lieb-Robinson velocity for the speed of information propagation $v_{LR}$ \cite{Lieb:1972wy}. We can obtain an open version of the drive $\tilde{U}^B_{\rm op}$ by cutting open the 1D edge and consistently excluding all terms that connect sites across the cut. Due to the Lieb-Robinson bound, this cannot affect the unitary evolution outside of a region $\Delta j\approx v_{LR}T$ near the left and right ends. For a long chain, with $L\gg v_{LR}T$, we expect the unitary far from the cut to act as before.

Now, assume that in the large length of chain from site $j=-M$ to site $j=M$, the action of the unitary is unaffected by the cut, as shown in Fig.~\ref{fig:edge_cartoon}(b). Since the charge in the bulk is transported uniformly by one lattice site through the action of the unitary, it follows from charge conservation that the charge initially in sites $\{-L, -L+1,\ldots,-M-1\}\cup\{M,M+1,\ldots L\}$ must equal the final charge in sites $\{-L, -L+1,\ldots,-M\}\cup\{M+1,M+2,\ldots L\}$. However, the available space for charge on the right is reduced by this evolution, while the space for charge on the left is increased. The only way that total charge can be conserved for \emph{any} initial charge configuration is if particles are transferred from the right edge to the left edge to address any imbalance. This distance can be made arbitrarily large by increasing the system size, which shows that in general, $\tilde{U}^B_{\rm op}$ must be nonlocal (or that $T$ must be infinite). We conclude that the anomalous action of $U^B_{\rm eff}$ cannot arise as result of a local 1D Hamiltonian $H(t)$ acting for a finite time.

For fermionic models, this argument shows that there is no local 1D unitary which brings the action of the open system to that of the closed system. Furthermore, for the doubled fermion drive, the above bosonic argument can be straightforwardly applied to demonstrate the anomalous nature of the edge unitary.

\textit{Exchange Models.}---
The models described above may be generalized straightforwardly to spin models or indeed any system where the on-site Hilbert spaces are equivalent. Instead of particle hops, the building blocks are now pairwise exchanges of local states. The exchange version of Eq.~\eqref{eq:2site_boson_unitary} is
\begin{equation}
U_{\br,\br'}^{\leftrightarrow}=\sum_{\alpha\neq \beta}\ket{\br,\beta}\otimes\ket{\br',\alpha}\bra{\br,\alpha}\otimes\bra{\br',\beta}+\delta_{\alpha\beta}\id_{\br\br'},
\end{equation}
where $\alpha,\beta\in\mathcal{H}_{\br}$ take values in the on-site Hilbert space. In the above, $\ket{\br,\alpha}$ indicates that the state at site $\br$ is $\alpha$.

In each step of the drive, the tensor product of the exchange operation is taken over one of the four sets of neighboring bonds shown in Fig.~\ref{fig:unitary_drive}(a), $U_j^{\leftrightarrow}=\bigotimes_{\br\in A}U_{\br,\br+\bb_j}^{\leftrightarrow}$, with the complete drive given by $U^{\leftrightarrow}=U_4^{\leftrightarrow}U_3^{\leftrightarrow}U_2^{\leftrightarrow}U_1^{\leftrightarrow}$. Each step consists of a product of local commuting terms, and so can be generated by a local Hamiltonian. The complete action of the drive may again be represented as in Fig.~\ref{fig:unitary_drive}(b), where the arrows now indicate the trajectory of a particular on-site state through the lattice. Acting on a product state, $U^{\leftrightarrow}$ permutes the on-site states through a cyclic permutation at the edge, an action that is also well defined for superposition states. This may be encapsulated in an effective edge unitary $U^{\leftrightarrow}_{\rm eff}$.

A natural setting for this type of anomalous drive is a lattice of spins. If the on-site Hilbert space corresponds to $\zbb_2$, then the model maps formally onto the hardcore boson model given previously. More general spin models may be mapped onto bosonic models that allow a different (but finite) number of particles per site. The Hamiltonians that generate these drives conserve the total boson number, and their edge action can be shown to be anomalous using the arguments given previously.

\textit{Stability of Edge Unitaries.}---
We now allow for the possibility of perturbations that do not conserve charge. In these cases, we can appeal to more general information theoretic ideas to show that the effective edge action of the Floquet drive is still anomalous. Roughly speaking, the anomalous edge drives have a chiral flow of information (and not just charge), which we will show cannot occur through a local 1D unitary evolution. 

As before, we begin by assuming that $U^{\leftrightarrow}_{\rm eff}$ may be generated by a local Hamiltonian $H(t)$, and so there is a maximum velocity $v_{LR}$ at which information can flow. We cut open the chain to obtain the putative open system unitary $\tilde{U}_{\rm op}^{\leftrightarrow}$, which should reproduce the permutation action in the bulk of the chain away from the cut.

\begin{figure}[t!]
\includegraphics[scale=0.27]{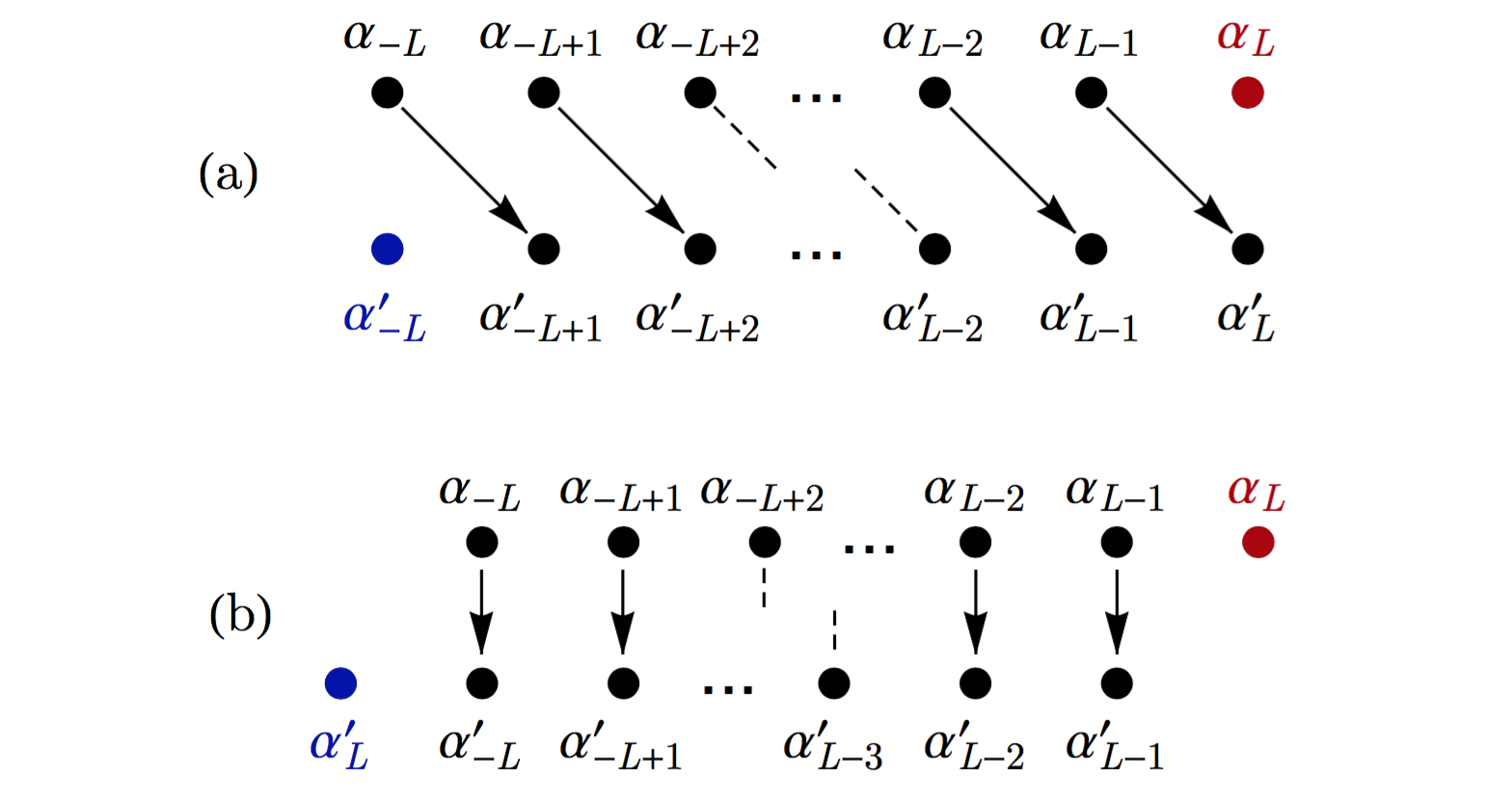}
\caption{(a) Permutation of on-site states under the action of $\tilde{U}_{\rm op}^{\leftrightarrow}$, assuming the 1D chain is cut between sites $L$ and $-L$. (b) Relabeling of lattice sites so that the permutation is diagonal away from the cut. \label{fig:infoedge}}
\end{figure}

For simplicity, we will assume that the edge region (of size $v_{LR}T$) consists of a single site on either side of the cut (for the more general case, see the Supplemental Material \cite{Note1}). With this setup, the action of $\tilde{U}_{\rm op}^{\leftrightarrow}$ on a many-body state is to translate the on-site states to the right by one lattice site, as shown in Fig.~\ref{fig:infoedge}(a). The unitary $\tilde{U}_{\rm op}^{\leftrightarrow}$ maps product states onto product states through the matrix elements $\ket{\alpha'_{-L},\alpha'_{-L+1},\ldots,\alpha'_{L}}=U^{\{\alpha'\}}_{\{\alpha\}}\ket{\alpha_{-L},\alpha_{-L+1},\ldots,\alpha_{L}}$, where $\ket{\{\alpha\}}$ and $\ket{\{\alpha'\}}$ are initial and final states, respectively. These matrix elements have the form 
\beq
U^{\{\alpha'\}}_{\{\alpha\}}=\prod_{j=-L}^{L-1}\delta_{\alpha^{\phantom{\prime}}_j,\alpha'_{j+1}}f\left(\{\alpha\},\{\alpha'\}\right),
\eeq
where the final factor describes the relation between states $\alpha_L$ and $\alpha'_{-L}$.

We now relabel the site indices in the final state through $j\to j'=j+1$ in the range $-L\leq j'\leq L$, which makes the permutation diagonal [see Fig.~\ref{fig:infoedge}(b)]. In this new basis, the matrix elements of the unitary are 
\beq
U^{\prime\{\alpha'\}}_{\{\alpha\}}=\delta_{\alpha^{\phantom{\prime}}_\bv,\alpha'_\bv} f\left(\alpha^{\phantom\prime}_\bv=\alpha'_\bv,\alpha^{\phantom{\prime}}_L,\alpha'_L\right),
\eeq
where we have used the shorthand notation $\bv$ to represent the sites from $-L$ to $L-1$. From the unitarity of $U^{\prime\{\alpha'\}}_{\{\alpha\}}$, it may be shown that the factor $f\left(\alpha_\bv,\alpha_L,\alpha'_L\right)$ is also unitary. The unitary evolution $\tilde{U}_{\rm op}^{\leftrightarrow}$ therefore relates state $\alpha'_L$ to $\alpha_L$ through an (unspecified) unitary operation.

Now, the unitary evolution as a whole preserves information. The diagonal factor in $U^{\prime\{\alpha'\}}_{\{\alpha\}}$ shows that information in the bulk is translated, while the remaining factor $f\left(\alpha_\bv,\alpha_L,\alpha'_L\right)$ shows that the information in state $\alpha_L$ is transferred to state $\alpha_L'$ unitarily. However, in the original basis, these states were separated by an arbitrarily large distance, and transferring information across this distance in a finite time would violate the Lieb-Robinson bound. In this way, the anomalous edge action of a general exchange model is robust and cannot be created or destroyed by a 1D unitary drive of the form $\tilde{U}_{\rm op}^{\leftrightarrow}$.

We can construct more general drives by stacking together several systems (and thereby acting on a tensor product Hilbert space) or by running several drives in sequence. Drives generated in this way are not necessarily independent, as we now show. 

The action of a general exchange drive can be characterized by a permutation of the form $\ket{\alpha_1,\alpha_2,\ldots,\alpha_{L}}\to\ket{\alpha_{L-p+1},\alpha_{L-p+2},\ldots,\alpha_{L-p}}$, which moves each state on the edge to the right by $p$ lattice sites. If the on-site Hilbert space has dimension $k$, then we write this right-moving permutation as $R(p,k)$. Left-moving permutations may similarly be written $L(p,k)$. 

We note that running the drive $R(p,k)$ $q$ times is equivalent to running the drive $R(qp,k)$ once. Secondly, we note that by grouping together the first $p$ lattice sites into a single effective site, the drive $R(p,k)$ is equivalent to the drive $R(1,k^p)$ \footnote{This regrouping is only formally possible if $L$ is a multiple of $p$. However, this makes no difference in the thermodynamic limit, and even for finite systems, the amount of information transferred is consistent with this expression.}. This regrouping of sites is equivalent to the stacking together of $p$ drives on different Hilbert spaces with dimension $k$. Stacking more general drives leads to the equivalence
\beq
R(p,k)\otimes R(p',k')\equiv R\left(1,k^p\left(k'\right)^{p'}\right)).
\eeq
In this way, any right-moving drive is equivalent to a drive $R(1,n)$, where $n$ is a positive integer. Using the same methods as above, it is straightforward to show that drives corresponding to different $n$ are inequivalent [i.e., $R(1,n)$ cannot be obtained from $R(1,n')$ through a local 1D unitary evolution for $n\neq n'$].

In the Supplemental Material \cite{Note1}, we show that by also including left-moving permutations, a generic permutation can be brought into the form $L(1,n')\otimes R (1,n)$, where $n$ and $n'$ are coprime integers.  A general exchange drive may therefore be characterized by a pair of integers, describing left- and right-moving components of the permutation. A trivial drive can be reduced to the form $n=1,n'=1$. Again using the methods above, all of these drives can be shown to be inequivalent.

In general, an effective edge unitary $U_{\rm eff}$ will not correspond to a pure exchange drive. From our results, it follows that any effective edge unitary that is equivalent to an exchange effective edge unitary can be characterized by coprime integers $n,n'$. We conjecture that this classification is also complete, i.e., that every effective edge unitary belongs to one of these equivalence classes.

\textit{Conclusions.}---
In summary, we have presented a many-body version of the anomalous Floquet drive of Ref.~\onlinecite{Rudner:2013bg}, which is applicable to both bosonic and fermionic systems. The action of the drive leads to the robust chiral propagation of charge at the boundary of an open system. Anomalous edge behavior arises more generally in exchange models, where spin states, for example, are swapped between Hilbert spaces on neighboring sites. Finite bounds on the propagation of information mean that classes of anomalous edge behavior are stable to all local perturbations. 

We showed that exchange drives may be uniquely characterized (up to equivalence) by a pair of coprime integers, and we conjectured that all effective edge unitaries are equivalent to one of these exchange drives. Using homotopy arguments \cite{Note1,Roy:2016ka}, these integers provide a topological classification of Floquet systems, including MBL phases and time crystals. Our work raises a number of interesting questions, which we hope will encourage further theoretical and experimental efforts. For instance, it would be interesting to study the interacting analogs and stability of other single-particle Floquet topological insulators.

\begin{acknowledgments}
We thank D.~Reiss, X.~Chen and the authors of Ref.~\cite{Po:2016iq} for useful discussions. R.~R. and F.~H. acknowledge support from the NSF under CAREER DMR-1455368 and the Alfred P. Sloan foundation.

\textit{Note added}---Recently, we became aware of Ref.~\cite{Po:2016iq}, which considers chiral Floquet phases in the context of MBL systems. Our results, while framed in a different setting, seem consistent with this work. Some differences are in the precise definition of effective edge unitaries and in the use of MBL for setting up the discussion. Reference~\cite{Po:2016iq} also suggests an experimental realisation for these systems and includes an explicit topological index, which we believe could also be used to classify our effective edge unitaries.
\end{acknowledgments}

%

\clearpage
\onecolumngrid

\setcounter{equation}{0}
\setcounter{figure}{0}
\setcounter{table}{0}
\setcounter{page}{1}
\makeatletter
\renewcommand{\theequation}{S\arabic{equation}}
\renewcommand{\thefigure}{S\arabic{figure}}

\begin{center}
\textbf{\large Supplemental Material}
\end{center}

\section{The Phase Space of Unitary Evolutions and Many-body Localization}
In the main text we focused on a specific class of drives known as unitary loops, which have the property that $U(T)=\id$ (in a closed system). We asserted that a topological classification of unitary loops in fact provides a much more extensive classification that may be applied to general time-dependent systems. Under the right circumstances, this classification also permits a well-defined notion of phase. In this section, we provide arguments to support these claims and make connections to many-body localization (MBL), a tool which is commonly used to study Floquet systems in the literature. Further details on the homotopy ideas introduced here may be found in Ref.~\onlinecite{Roy:2016ka}.

A generic unitary evolution $U(t)$ may be thought of as a path in the space of unitary operators, $\mathcal{S}_0$, which begins at the identity and which ends at some point $P$. Two unitary evolutions are then equivalent, up to homotopy, if one can be deformed continuously into the other without leaving the space $\mathcal{S}_0$. On its own, this unconstrained definition of homotopy is not very useful, since any unitary evolution may be smoothly deformed into any other without leaving the space $\mathcal{S}_0$. To provide a meaningful classification, we should consider only unitary evolutions which have endpoints restricted to lie in some region $W$. For example, $W$ may be a single point in the space $\mathcal{S}_0$, in which case two unitary evolutions which end at this point, $U_1(t)$ and $U_2(t)$, have a natural relative classification given by the topological order of the loop, $U_2^{-1}\circ U_1$, which consists of the evolution $U_1(t)$ followed by the evolution $U_2^{-1}(t)$. The order of this loop is given by a pair of coprime integers as described in the main text.

A more useful choice for $W$ is the set of endpoints that take the form $e^{-iH_F}$, where $H_F$ is some local Floquet Hamiltonian. As argued in Ref.~\onlinecite{Roy:2016ka}, we may define the `standard' path to this endpoint as the constant evolution due to Hamiltonian $H_F$, writing $U_{H_F}(t)=e^{-iH_Ft}$. For a unitary evolution $U(t)$, this permits a notion of absolute loop order by considering the loop $U_{H_F}^{-1}\circ U$.

These considerations then naturally also apply to endpoints of the form $e^{-iH_{\rm MBL}}$, where $H_{\rm MBL}$ is an MBL Hamiltonian. In an MBL system, weak disorder breaks ergodicity and permits a description of many-body states in terms of local, conserved quantities known as l-bits (see Ref.~\onlinecite{Nandkishore:2015kt} for a review). In this situation, the notion of quantum phase may be applied to the complete spectrum of eigenstates (or quasieigenstates), and becomes well-defined even at finite temperature \cite{Bahri:2015ib,Potter:2015vna,Slagle:2015uo,Chandran:2014dk,Bauer:2013jw,Huse:2013bw}. Moreover, since a driving Hamiltonian transfers energy to the system, MBL is believed to be required to prevent heating to infinite temperature \cite{Zhang:2016hy,Zhang:2016vt,Khemani:2016gd,Abanin:2015bc,Lazarides:2015jd,Ponte:2015dc,Ponte:2015hm,Abanin:2016eva,Lazarides:2014ie,DAlessio:2013fv}, and may also be needed for the observation of topological signatures such as persistent edge modes \cite{Chandran:2014dk,Bahri:2015ib}. MBL unitaries are therefore particularly robust to perturbations, although we note that there is some doubt about the existence of MBL in dimensions greater than one \cite{DeRoeck:2016us}. Nevertheless, if we choose the region $W$ to be endpoints of this form, with an additional plausible assumption about the absence of loops in this region, the classification decomposes naturally into a classification of endpoint MBL Hamiltonians, $\{H_{\rm MBL}\}$, and a classification of possible loop orders. This order is then robust to perturbations (subject to caveats about the existence of MBL) and provides a well-defined notion of phase. In this way, the nontrivial loops described in the main text provide a classification of Floquet MBL phases.

The loop order defined in this article, however, is also applicable to more general choices for $W$, including endpoint unitaries that are only partially many-body localized, and endpoint unitaries that describe time crystals (also known as $\pi$-spin glasses) \cite{Khemani:2016gd,vonKeyserlingk:2016bq,Else:2016gf,vonKeyserlingk:2016ev,Yao:2017bu}. A complete discussion of these endpoint choices may be found in Ref.~\onlinecite{Roy:2016ka}.

\section{Proof of the Existence of an Effective Edge Unitary}
In this section, we prove that a unitary evolution that behaves as the identity in a closed system always yields a well-defined effective edge unitary in the open system. The effective edge unitary will be shown to be localized near the boundary of the open system. We will implicitly work in two dimensions, assuming the closed system is a torus and the open system has been cut open to give a cylinder.

We write the time-dependent Hamiltonian that generates the unitary loop as $H(t)$. Since this must be local, we can consistently define the Hamiltonian of the open system by excluding the terms that connect parts of the system on either side of the cut. Specifically, we write
\beq
H_{\rm op}(t)&=&H_{\rm cl}(t)+H_{\rm e}(t),
\eeq
where $H_{\rm cl}(t)$ is the Hamiltonian of the closed system, $H_{\rm op}(t)$ is the Hamiltonian of the open system, and $H_{\rm e}(t)$ includes (the negative of) the edge terms that connect sites either side of the cut. 

We define unitary evolutions of the closed and open systems over the time interval $[t_1,t_2]$ in the usual manner,
\beq
U_{\rm cl/op}(t_2,t_1)&=&\mathcal{T}\exp\left[-i\int_{t_1}^{t_2}H_{\rm cl/op}(t')\,\dd t'\right],
\eeq
where $\mathcal{T}$ is the time-ordering operator. We now define a modified edge Hamiltonian through
\beq
H_{\rm e}'(t)&=&U_{\rm cl}\left(T,t\right)H_{\rm e}(t)\left[U_{\rm cl}\left(T,t\right)\right]^{-1}.\label{eq:modified_edge_h}
\eeq
Since all parts of the Hamiltonian are local, the modified Hamiltonian $H_{\rm e}'(t)$ is also local: conjugation with $U_{\rm cl}\left(T,t\right)$ can at most increase the range of Hamiltonian $H_{\rm e}(t)$ by a length $v_{LR}(T-t)$, where $v_{LR}$ is the (maximum) Lieb-Robinson velocity of $H_{\rm cl}(t)$.

Defining the infinitesimal time interval $\Delta t\equiv \lim_{N\to\infty}T/N$, we write the open system unitary as the infinite product
\beq
U_{\rm op}(T)&=&e^{-iH_{\rm cl}(T)\Delta t}e^{-iH_{\rm e}(T)\Delta t}e^{-iH_{\rm cl}(T-\Delta t)\Delta t}e^{-iH_{\rm e}(T-\Delta t)\Delta t}\ldots e^{-iH_{\rm cl}(\Delta t)\Delta t}e^{-iH_{\rm e}(\Delta t)\Delta t}.\label{eq:open_unitary}
\eeq
Then, recalling Eq.~\eqref{eq:modified_edge_h}, we further rewrite the open system unitary in terms of the modified edge Hamiltonian as
\beq
U_{\rm op}(T)&=&e^{-iH'_{\rm e}(T)\Delta t}e^{-iH'_{\rm e}(T-\Delta t)\Delta t}\ldots e^{-iH'_{\rm e}(\Delta t)\Delta t}U_{\rm cl}(T),\label{eq:open_unitary_modified}
\eeq
where in the final factor $U_{\rm cl}(T)$, we have already already taken the infinitesimal limit to sum the terms involving $H_{\rm cl}(t)$. However, by definition, the closed system unitary operator is just the identity, and so we identify
\beq
U_{\rm op}(T)=\mathcal{T}\exp\left[-i\int_{0}^{T}H'_{\rm e}(t')\,\dd t'\right]\equiv U_{\rm eff}.
\eeq
The Hamiltonian $H'_{\rm e}(t)$ is localized near the boundary of the open system, and so $U_{\rm eff}$ is an effective edge operator that also acts only in the vicinity of the boundary. Other than locality, we have not assumed any properties of the Hamiltonian $H(t)$, and so the effective edge unitary is well defined for bosonic, fermionic and spin systems. 

Finally, $U_{\rm op}(T)$ cannot entangle the two ends of the cylinder since they are separated by an arbitrarily large distance. The effective unitary must therefore factorize as $U_{\rm eff}=U^L_{\rm eff}U^R_{\rm eff}$, where $L$ and $R$ label the two ends of the cylinder.
\section{Information Flow with a Finite Edge Region}
In the main text, we showed that the anomalous edge action of an exchange model was robust to local evolutions at the edge, by cutting open the 1D edge and assuming only a single site each side of the cut was affected. More generally, there will be a finite edge region affecting $M\approx v_{LR}T$ sites either side of the cut. In this case, since the information in the bulk is translated, the information initially in sites $\{-L,-L+1,\ldots,-M-1\}\cup\{M,M+1,\ldots,L\}$ must be related to the information that ends up in sites $\{-L,-L+1,\ldots,-M\}\cup\{M+1,M+2,\ldots,L\}$. (This is similar to the charge conservation argument given in the main text). From the unitarity of the complete evolution, the information in these edge regions must be related through a unitary transformation $U^{L'R'}_{LR}$, where $L,R$ and $L',R'$ label the initial and final edge regions, respectively. However, this unitary cannot factorize into the local form $U_{L}^{L'}U^{R'}_{R}$ because $\mathrm{dim}(R')<\mathrm{dim}(R)$ and $\mathrm{dim}(L')>\mathrm{dim}(L)$. The only way that information can be preserved is if some is transferred from $R$ to $L'$, which would violate the Lieb-Robinson bound.
\section{Example of a Unitary Loop with Trivial Effective Edge Unitary}
\begin{figure}[h]
\includegraphics[scale=0.4]{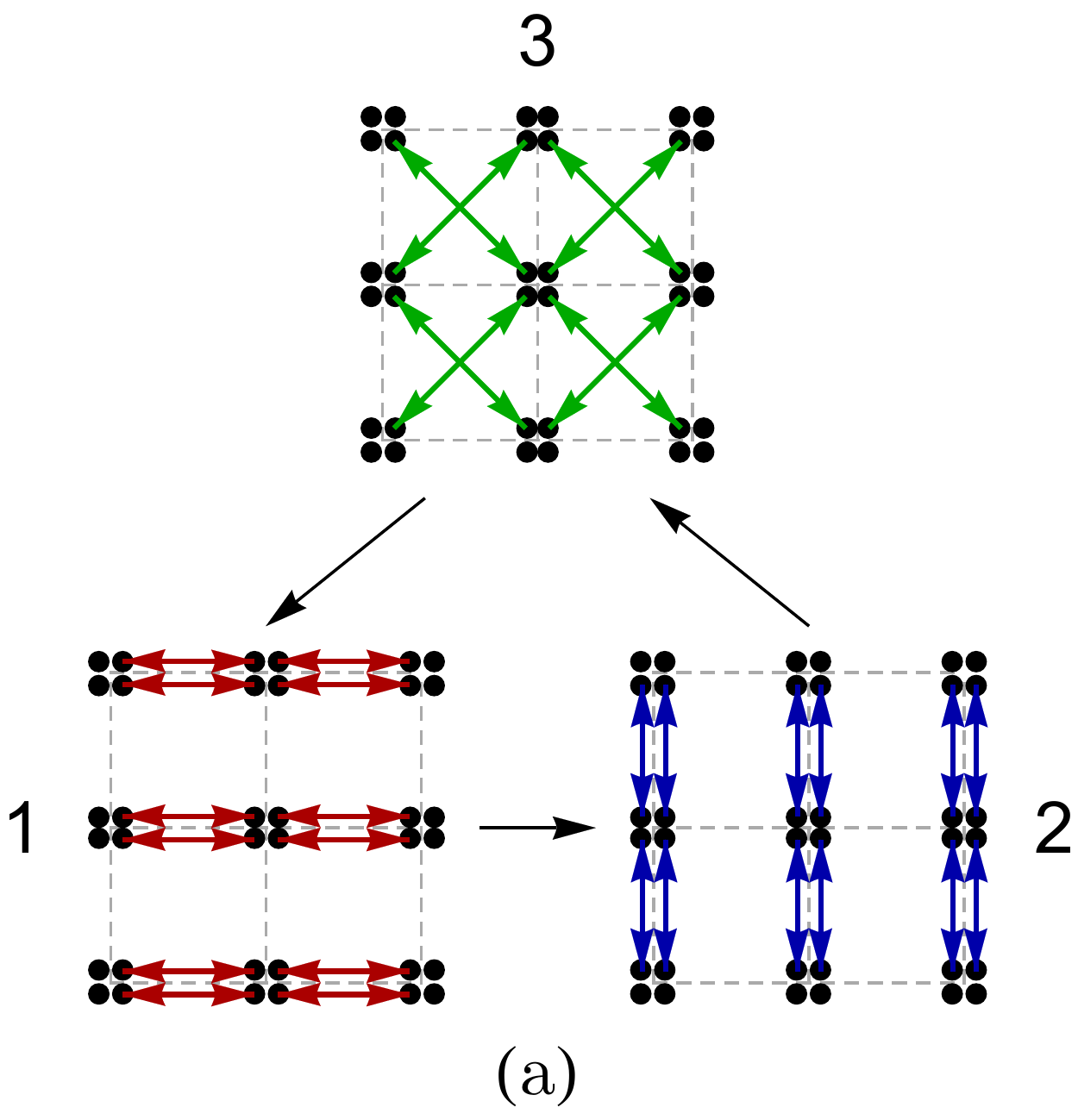}\hspace{5mm}
\includegraphics[scale=0.35]{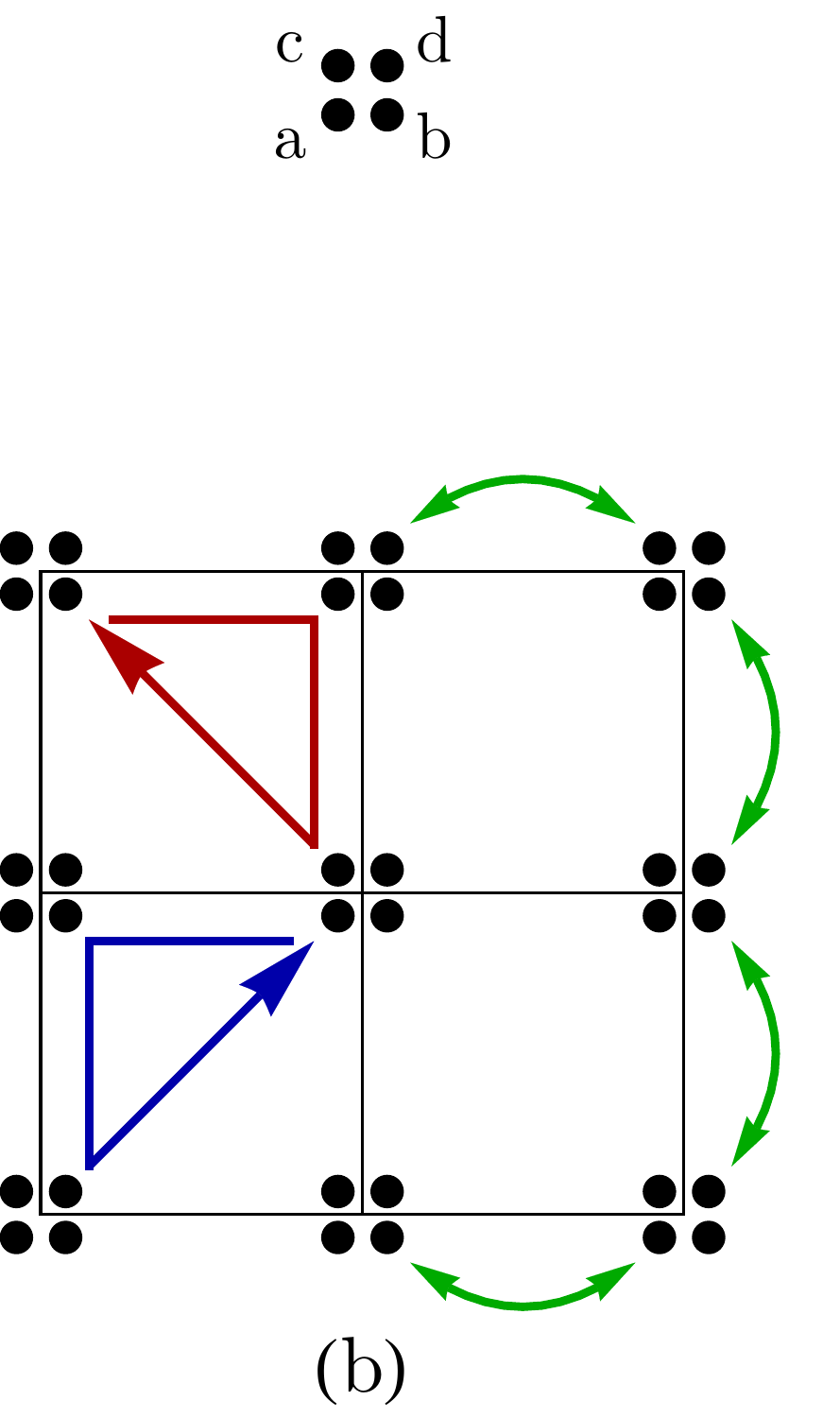}
\caption{(a) Example of a unitary loop that generates a trivial effective edge unitary that is not the identity. On-site degrees of freedom have been drawn spatially separated for clarity. (b) Labeling of the on-site degrees of freedom and representation of the action of this drive. See main text for details. \label{fig:trivial_loop}}
\end{figure}
In this section, we give an explicit example of a unitary loop that gives an effective edge unitary which, while not equal to the identity, is in the trivial class of permutations. This means that the effective edge unitary may be reduced to the identity through the action of local unitaries.

We consider a square lattice with four sublattice degrees of freedom per site, as shown in Fig.~\ref{fig:trivial_loop}. Writing the wavefunction corresponding to state $\alpha$ on site $\br$ and sublattice $a$ as $\ket{\br,a,\alpha}$, a general exchange operation is written
\beq
U_{\br ,a;\br' ,a'}^{\leftrightarrow}&=&\sum_{\alpha\neq \beta}\ket{\br,a,\beta}\otimes\ket{\br',a',\alpha}\bra{\br,a,\alpha}\otimes\bra{\br',a',\beta}+\delta_{\alpha\beta}\id_{\br,a;\br',a'}.
\eeq
The drive then consists of the three steps,
\beq
U_1&=&\prod_{\br}U_{\br ,b;\br+\bx, a}^{\leftrightarrow}U_{\br ,d;\br+\bx, c}^{\leftrightarrow}\nonumber\\
U_2&=&\prod_{\br}U_{\br ,c;\br+\by, a}^{\leftrightarrow}U_{\br ,d;\br+\by ,b}^{\leftrightarrow}\\
U_3&=&\prod_{\br}U_{\br ,d;\br+\bx+\by, a}^{\leftrightarrow}U_{\br ,b;\br+\bx-\by ,c}^{\leftrightarrow}.\nonumber
\eeq
The overall action of this drive is represented pictorially in Fig.~\ref{fig:trivial_loop}(b): in the bulk, states return to their initial locations after three steps, while on the edge, some neighboring states are swapped. The effective edge unitary is therefore a commuting product of pairwise exchanges on the edge, which may be generated by a local Hamiltonian $H(t)$. A Hamiltonian that undoes these local exchanges will reduce the effective edge unitary to the identity.

\section{Further Details on The Classification of Permutations}
\begin{figure}[h]
\includegraphics[scale=0.4]{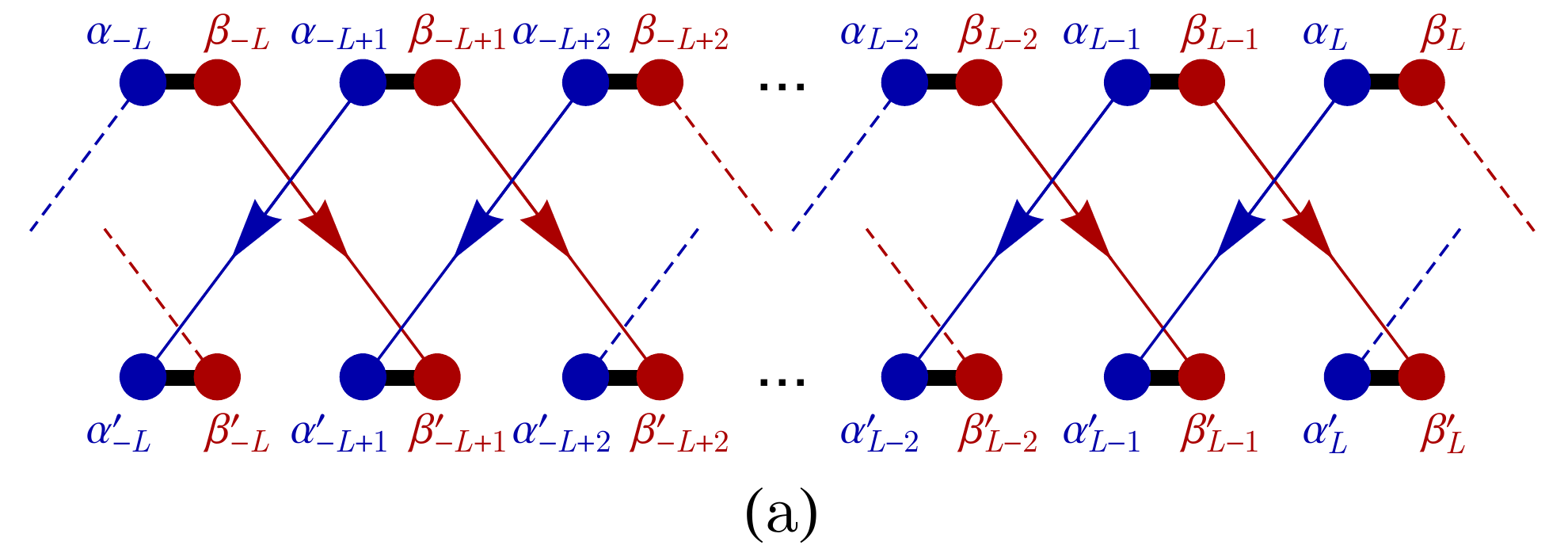}\hspace{5mm}
\includegraphics[scale=0.4]{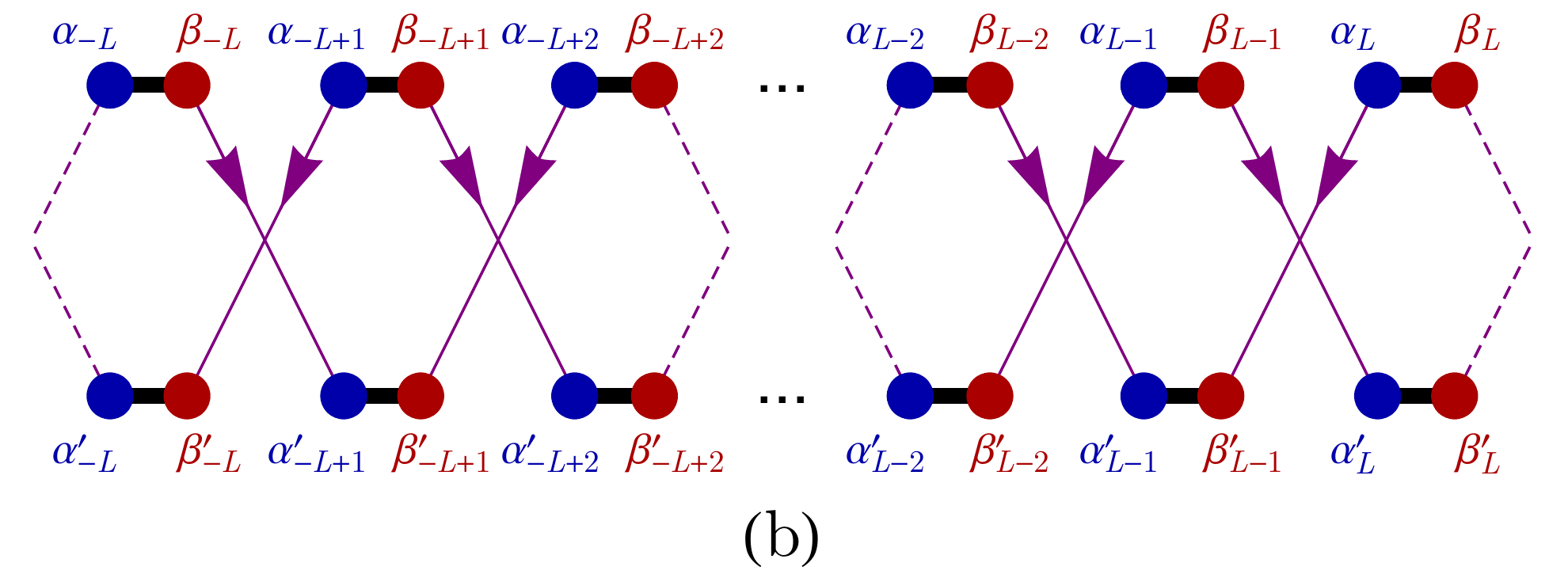}
\caption{(a) Representation of the composite drive $L(1,k)\otimes R(1,k)$, with two copies of the Hilbert space $\mathcal{H}_\br$ per site. States from one on-site Hilbert space are translated to the left by the action of the drive, while states from the other on-site Hilbert space are translated to the right. (b) Representation of the composite drive $L(1,k)\otimes R(1,k)$ after acting with a product of local unitaries that interchanges the two on-site Hilbert spaces in the final state. The action of the drive is now a product of pairwise swaps. See main text for details. \label{fig:LRedge}}
\end{figure}

In the main text, we showed that any combination of right-moving permutations can be cast into the form $R(1,n)$, where $n$ is a positive integer. We now also consider left-moving permutations. It is clear that $R(p,k)$ followed by $L(q,k)$, with $q<r$ is equivalent to $R(p-q,k)$. It can also be shown that the stacking $L(p,k)\otimes R(p,k)$ is trivial. To argue this, we will show that $L(1,k)\otimes R(1,k)$ is trivial, from which the more general result follows.

The product drive $L(1,k)\otimes R(1,k)$ acts on the on-site tensor product Hilbert space $\mathcal{H}_\br\otimes\mathcal{H}_\br$, represented by the blue and red circles in Fig.~\ref{fig:LRedge}. Under the action of the drive, states from one Hilbert space move one site to the right, while states on the other Hilbert space move one site to the left. The characteristic permutation nature of the drive is preserved under the action of local unitaries, however, which may mix together the on-site Hilbert spaces and act on clusters of nearby sites.

In this case, we imagine acting after the drive with a product of local unitaries, each of which interchanges the two states on a given site. If the two on-site Hilbert spaces are labelled $a$ and $b$, then the required local unitary is
\begin{equation}
U_{\br,a;\br,b}^{\leftrightarrow}=\sum_{\alpha\neq \beta}\ket{\br,a,\beta}\otimes\ket{\br,b,\alpha}\bra{\br,a,\alpha}\otimes\bra{\br,b,\beta}+\delta_{\alpha\beta}\id_{\br,a;\br,b}.
\end{equation}
Local unitaries of this form from each site commute, and so the complete product may be generated by a local Hamiltonian acting for a finite time. The action of this local unitary changes the effect of the drive to a product of pairwise swaps of states between neighboring sites, as shown in Fig.~\ref{fig:LRedge}(b). 

Once in this form, it is clear that the action of the unitary is local and that it belongs to the trivial permutation class. Explicitly, the transformed unitary may now be written
\beq
U^{L(1,k)\otimes R(1,k)}&=&\bigotimes_\br U^{\leftrightarrow}_{\br,b;\br+\bx,a},
\eeq
where $U^{\leftrightarrow}_{\br,b;\br+\bx,a}$ is the local pairwise exchange between the state in Hilbert space $b$ on site $\br$, and the state in Hilbert space $a$ on site $\br+\bx$. Its action may be reduced to the identity by the inverse operation, which is a local, finite-time unitary. 

Thus, we see that $L(1,k)\otimes R(1,k)$ is trivial, and it follows from this relation that 
\beq
L(1,nk')\otimes R(1,nk)\equiv L(1,k')\otimes R(1,k).
\eeq 
By gathering together left- and right-moving parts of the permutation, and then canceling common integer factors, a general composition of drives may be written $L(p',k')\otimes R(p,k)\equiv L(1,\left(k'\right)^{p'})\otimes R\left(1,k^p\right)$.

\end{document}